\documentclass[11pt]{article}
\usepackage[utf8]{inputenc}
\usepackage{graphicx}
\usepackage{amsmath, amssymb}
\usepackage{authblk}  
\usepackage{cite}     
\usepackage{geometry} 
\usepackage{caption}
\usepackage{hyperref} 

\title{Moth’s eye-inspired perfectly vertical subwavelength grating coupler for silicon photonics}

\author[1]{Ivan A. Kazakov}
\author[1]{Ilona  Popova}
\author[1]{Arkady Shipulin}

\affil[1]{Photonic Integration Research Lab, Skolkovo Institute of Science and Technology, Moscow, Russia}

\date{April 25, 2026}

\begin{document}

\maketitle

\begin{abstract}
We propose a novel bio-inspired design principle for the perfectly vertical grating coupler. The main idea of our design is to introduce anisotropy to the grating stripe to direct the light to one side of the grating. This grating design is easy to manufacture, only requiring a single etching step, and it is designed to efficiently couple vertically incident light. This makes it a good candidate for heterogeneous integration of light sources, especially VCSELs, on chip for applications in classical and quantum communications, LIDARs, sensing systems, and others. The grating coupler was designed for the SOI material platform with a central wavelength of 1550 nm. We obtained the efficiency of in-coupling from the  SMF-28 fiber of 41\% at vertical incidence and unidirectionality of over 10 dB, with a bandwidth of 50 nm at a 1 dB level in simulation. Experimental measurements confirmed unidirectionality, with observed unidirectionality of $12.80 \pm 0.02$ dB and a single-coupler insertion loss of $8.35 \pm 0.02$ dB around 1528 nm.
\end{abstract}

\noindent\textbf{Keywords:} silicon photonics, grating couplers, vertical coupling, anisotropic coupling, subwavelength structures


\section{Introduction}
The problem of coupling light to photonic integrated circuits (PICs) is essential for the integrated photonics technology. There are various ways of coupling light to PICs, which can be divided into two main branches: edge coupling (EC) and grating coupling (GC). Each of them have their own pros and cons, and are used in different applications \cite{marchetti2019coupling, cheng2020grating, vermeulen2018optical}. Currently, the most popular are edge couplers (ECs) due to their wide bandwidth, low loss, polarization insensitivity, and simple packaging. But there are a set of problems that can not be solved with them; for example, grating couplers (GCs) are essential for phased array antennas, wafer-level testing, and some experimental measurements. The main problem we try to solve with our design is the heterogeneous integration of VCSELs with PICs. VCSELs are the most energy- and cost-efficient light sources, and their integration to silicon PICs will be preferable for a lot of applications, such as LiDARs, classical and quantum communications, and sensing. Currently, there are several ways and attempts for coupling VCSELs to silicon on insulator (SOI) PICs \cite{lu2016flip}.

\begin{figure}
  \center{\includegraphics[width=13cm]{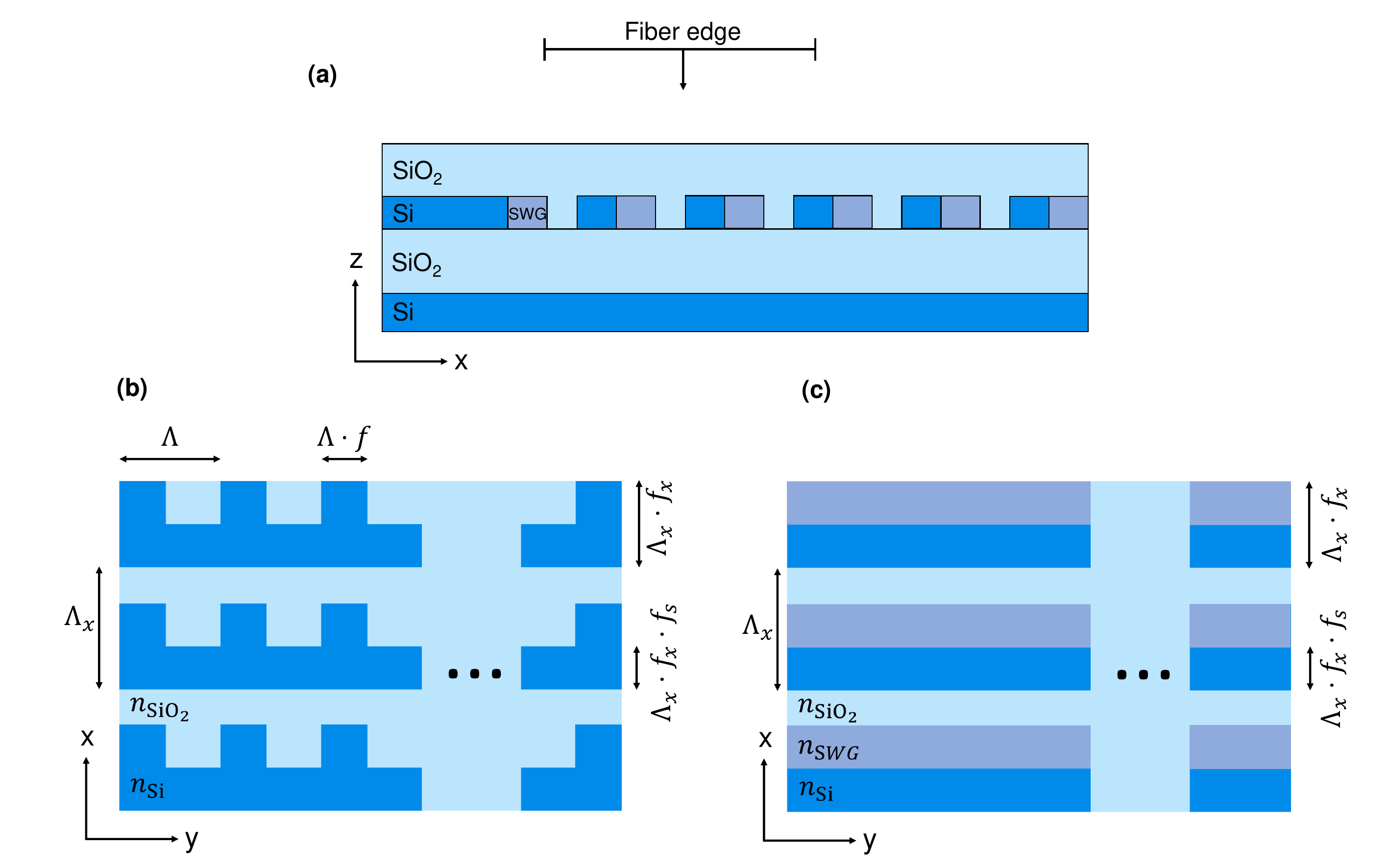}}
  \caption{(a) Side view of the proposed grating, which includes a sub-wavelength grating (SWG) region. (b) Top view of the proposed grating, showing the actual geometry of the grating, (c) top view showing the effective grating geometry, with $n_{SWG}$ obtained from eq. \eqref{eq:ryt12}}
  \label{fig:schematic}
\end{figure}

We suggest a new approach for designing a perfectly vertical grating coupler based on several assumptions: the grating should be made in one etching step for fabrication simplicity,  be unidirectional to avoid problems related to interference, which appear in bidirectional couplers, and have high efficiency of coupling vertically incident light.

An important concept used in our approach is the subwavelength grating (SWG), which allows to engineer the effective refractive index of the grating stripes, thus extending the variety of refractive indices available for design \cite{halir2015waveguide, cheben2018subwavelength}. For our GC, we have chosen the 220 nm SOI material platform as the most popular one for creating SWG structures due to high index contrast and the possibility of manufacturing devices with high resolution.  SWG is a periodic structure with period $\Lambda$ made of two materials with different refractive indices, in our case $n_{\textrm{Si}}$ and $n_{\textrm{SiO$_{2}$}}$. The effective refractive index of a subwavelength grating can be calculated using the effective medium theory (EMT), the idea of which is to decompose the effective index into a Taylor series with respect to the small parameter, which is period to wavelength ratio R (Eq. \ref{eq:r}).

\begin{equation}
   \label{eq:r} R = \frac{n_{eff} \Lambda}{\lambda}
\end{equation}

Here, $n_{\mathrm{eff}}$ denotes the effective index of the uniform waveguide, and in our case, it is equal to $n_{\textrm{Si}}$ because the y-dimension of our device is much larger than the wavelength.

If $R << 1$, which corresponds to the deep-subwavelength regime, SWG can be described by the zero-order effective medium theory (EMT), developed by the Rytov \cite{rytov1956electromagnetic}. Formulas for TE and TM polarisation are the following:

\begin{eqnarray}
\label{eq:ryt1} \frac{1}{n^{(0)}_{TE}} = \left( \frac{(1 - f_y)}{n_{SiO_2}^2} + \frac{f_y}{n_{Si}^2} \right)^{\frac{1}{2}} \\
\label{eq:ryt2}  n_{TM}^{(0)} = \left(n_{Si}^2 f_y + n_{SiO_2}^2 (1 - f_y) \right)^{\frac{1}{2}}
\end{eqnarray}

The condition $R << 1$ is unachievable with current technology, so for cases where $R<1$, a second-order approximation can be used \cite{lalanne1996effective, lalanne1998high}. This approximation is in good coincidence with experimental results. Equations are presented below:

\begin{eqnarray}
\label{eq:ryt12} n^{(2)}_{TE} = n^{(0)}_{TE} \left( 1 + \frac{\pi^2}{3} R^2 f_y^2 (1 - f_y)^2 (n_{SiO_2}^2 - n_{Si}^2 )^2 \left( \frac{n_{TM}^{(0)}}{n_{effTE}} \right) ^ 2 \left( \frac{n_{TE} ^ {(0)}}{n_{SiO_2} n_{Si}} \right) ^ 4 \right) ^ {\frac{1}{2}}\\
\label{eq:ryt22}  n^{(2)}_{TM} = n^{(0)}_{TM} \left( 1 + \frac{\pi^2}{3} R^2 f_y^2 (1-f_y)^2 \left( \frac{n_{SiO_2}^2 - n_{Si}^2}{n_{effTM n_{TM}^{(0)}} } \right) ^ 2 \right) ^ \frac{1}{2}
\end{eqnarray}

\begin{figure}
  \center{\includegraphics[width=10cm]{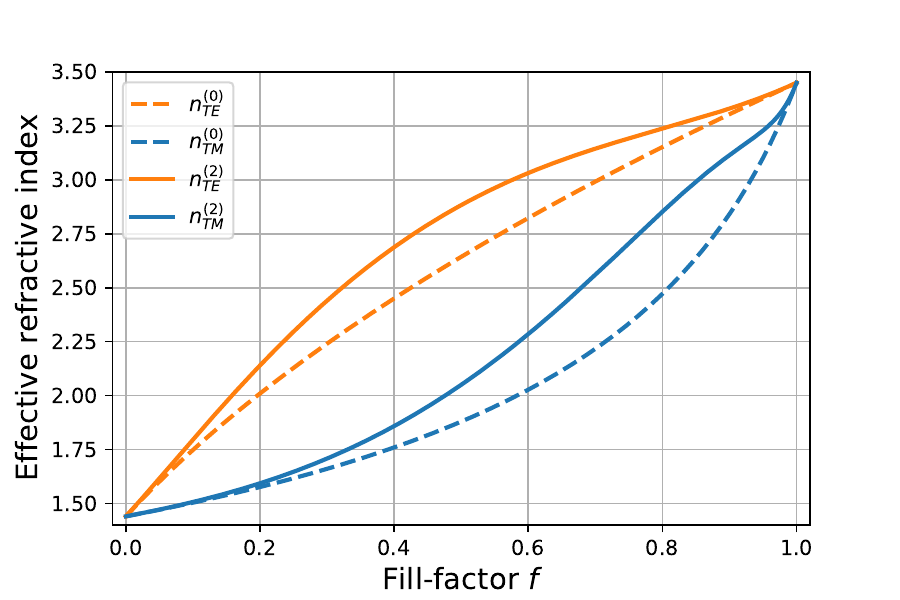}}
  \caption{Approximated refractive index of the sub-wavelength grating for different fill-factors, calculated by EMT with
zeroth-order and second-order approximation ($\Lambda = 400$ nm, $\lambda = 1550$ nm).}
  \label{fig:neff}
\end{figure}

The behavior of zeroth- and second-order effective indices as a function of fill-factor $f$ for the period $\Lambda = 400$ nm is shown in Figure \ref{fig:neff}. Sub-wavelength gratings can be used for refractive index engineering, and they allow to achieve any refractive index value between $n_{\textrm{SiO$_{2}$}}$ and $n_{\textrm{Si}}$ by choosing the appropriate value of fill-factor $f$. Effective index values retrieved from equations \eqref{eq:ryt12} and  \eqref{eq:ryt22} can be used for 2D-FDTD simulations in the plane of incidence.

Our design idea is based on breaking the symmetry of the grating stripe in order to break the symmetry of light propagation. Guiding light to one side is provided by the fact that one side of the grating stripe is flat and reflective, and the other side is covered with 'teeth' which act as an antireflective surface, so-called Moth's Eye metasurface \cite{han2016antireflective}. Therefore, light is absorbed from the patterned side of the stripe, while it is on the surface and therefore guided to this side. From another point of view, the patterned region is a sub-wavelength grating, which has its own effective refractive index; this approach allows for avoiding multi-level etching by patterning the waveguide in the transverse direction to create metamaterials with different effective refractive indices.

Further, we explain our simulation setup and numerical simulation and optimization procedure, and finally cover the question of comparison of our solution with others and ways of further development of our approach for GC design.

\section{Methods}
\begin{figure}[ht]
  \centering\includegraphics[width=10cm]{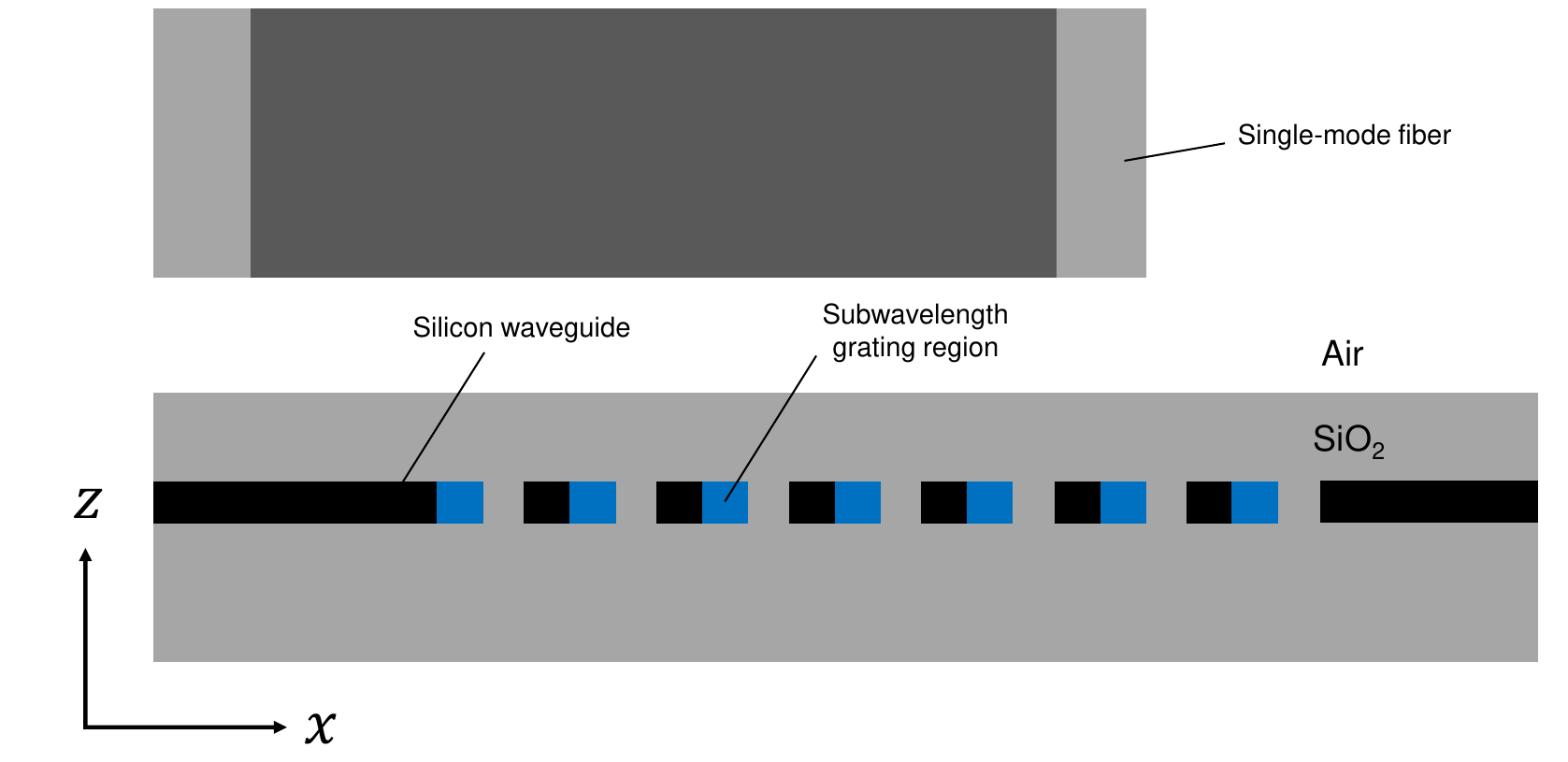}
  \caption{Cross section of the FDTD simulation setup}
  \label{fig:cross}
\end{figure}

To calculate coupling efficiency, we perform three-dimensional finite-difference time-domain (3D-FDTD) simulations using the open-source software MEEP \cite{oskooi2010meep}. The simulation setup is shown schematically in Figure \ref{fig:cross}. The grating is illuminated by a short pulse of light of the fundamental mode of a standard single-mode optical fiber. Simulation was run until the electric field had sufficiently decayed. After the simulation is finished, we calculate the Fourier transform of the electric field. Then, the Fourier-transformed electric field at the fiber cross-section is decomposed into fiber eigenmodes $E_{n}$ as:

\begin{equation}
\label{eq:fiber-field}
E^{fiber}(x,y,\omega)=\sum_{n}\alpha_{n}^{+}(\omega)E_{n}^{+}(x,y,\omega)+\alpha_{n}^{-}(\omega)E_{n}^{-}(x,y,\omega)
\end{equation}

And the electric field at the waveguide cross-section is decomposed into waveguide eigenmodes $\mathcal{E}_{n}$ as:

\begin{equation}
\label{eq:wg-field}
E^{wg}(y,z,\omega)=\sum_{n}\beta_{n}^{+}(\omega)\mathcal{E}_{n}^{+}(y,z,\omega)+\beta_{n}^{-}(\omega)\mathcal{E}_{n}^{-}(y,z,\omega)
\end{equation}

where the "+" and "-" superscripts correspond to forward- and backward-traveling waves, respectively, and $n=0$ corresponds to the fundamental $E_{y}$-polarized (TM) mode of both fiber and waveguide. The efficiency as a function of frequency is then calculated as:

\begin{equation}
\label{eq:ef_vs_freq}
\eta(\omega)=\frac{|\beta_{0}^{+}(\omega)|^{2}}{|\alpha_{0}^{+}(\omega)|^{2}}
\end{equation}

Using this definition of efficiency, i.e., separating forward- and backward-traveling waves, makes sure that light reflected from the grating back into the fiber does not affect the numerical results. 

As shown in Figure \ref{fig:schematic}, there are a total of five parameters to optimize: $f$, $f_{x}$, $f_{s}$, $\Lambda$ and $\Lambda_{x}$. We fixed the grating width to 12 $\mu$m and the grating length to 12 periods. Then, to get decent initial guesses for the other parameters, we ran a two-dimensional optimization, where the SWG region was represented by a region with refractive index $n_{SWG}$, which was calculated according to Eq. \eqref{eq:ryt12}. The obtained parameters were used as initial guesses for the 3D simulation, where we tried using both gradient-based and gradient-free optimization methods. Three-dimensional simulations were run on 16 computer cores with a grid size of 40 points per $\mu$m. The value of coupling efficiency $\eta(\omega_{0})$ at $\lambda_{0} = 1550$ nm was used as the figure of merit.

\section{Results}

After running the optimization procedure until convergence, we obtain a coupling efficiency of 41\% at 1550 nm. The values of corresponding grating parameters are given in Table \ref{tab:results_param}. 

\begin{table}[]
    \centering
    \begin{tabular}{c|c|c|c|c|c}
        Parameter & $f$ & $f_{x}$ & $f_{s}$ & $\Lambda$ & $\Lambda_{x}$ \\
        \hline
        Value & 0.50 & 0.72 & 0.48 & 470 nm & 750 nm 

    \end{tabular}
    \caption{Optimized grating parameters}
    \label{tab:results_param}
\end{table}

\begin{figure}[ht]
\centering\includegraphics[width=10cm]{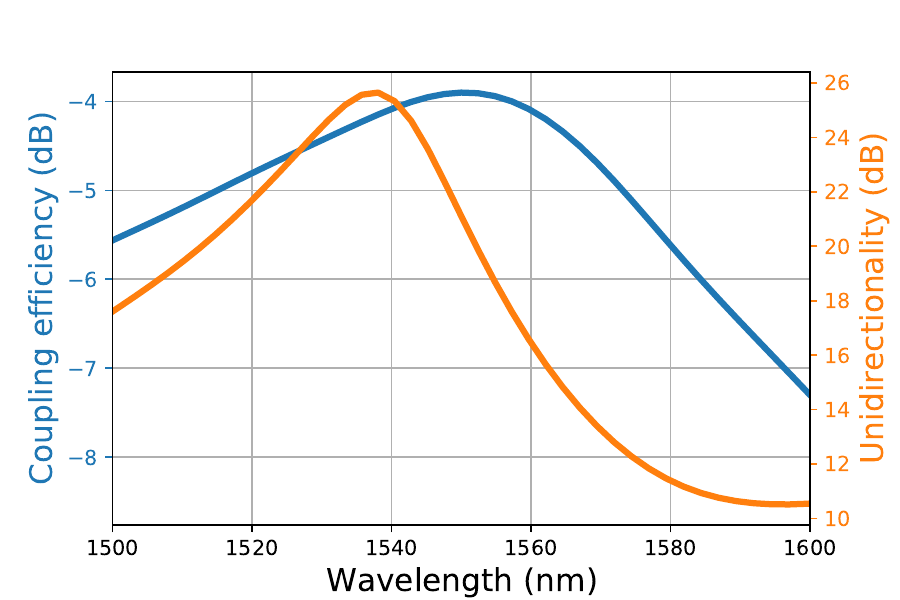}
\caption{Coupling efficiency (blue, left axis) and unidirectionality (orange, right axis) as a function of wavelength for the optimized grating}
\label{fig:eff_vs_wl}
\end{figure}

Since our figure of merit was only optimized for a single wavelength, we then calculated the coupling efficiency in a broad spectral range to make sure that the designed grating has a sufficiently high bandwidth. The corresponding plot is shown in Fig. \ref{fig:eff_vs_wl}, which shows that the coupling efficiency is higher than -5 dB in the entire C-band, with a maximum of 41\% at 1550 nm, and the -1 dB bandwidth is around 50 nm. Unidirectionality, which is defined as the difference (in dB) between the coupling efficiency to the right and to the left waveguide, is equal to 20 dB at 1550 nm, which means that less than 1 \% of the light is coupled into the other waveguide.

\begin{figure}[ht]
\centering\includegraphics[width=10cm]{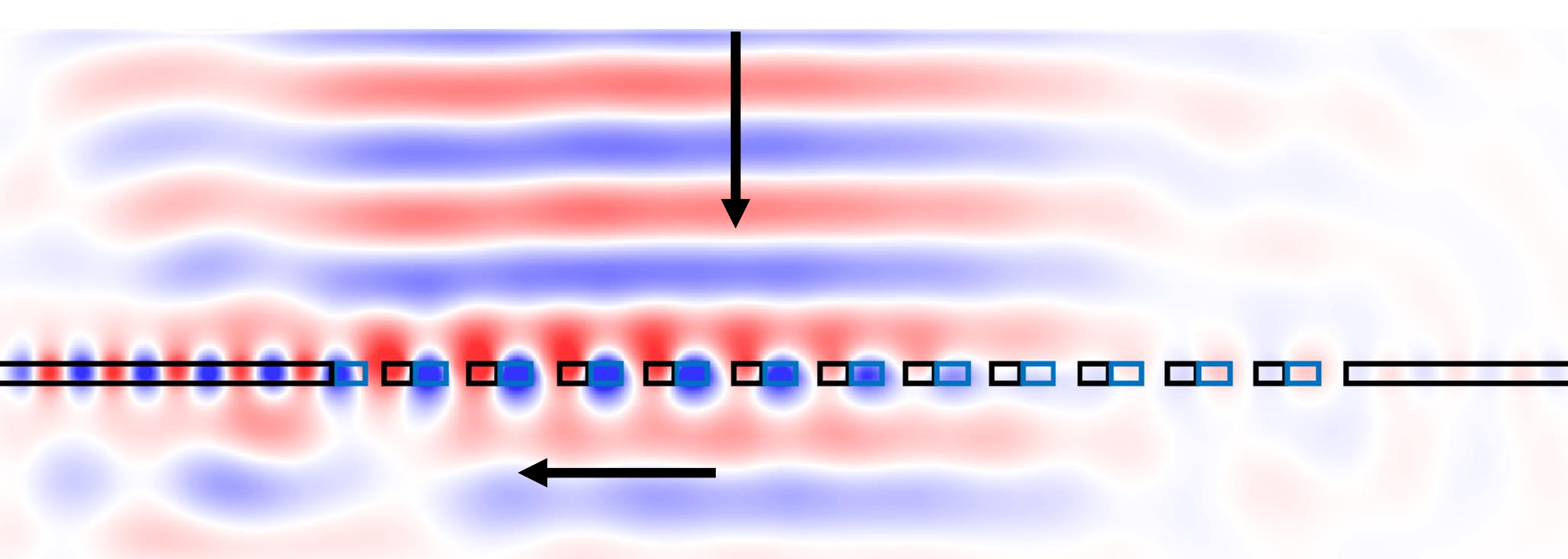}
\caption{Distribution of $E_{y}$ for the optimized grating. Black lines indicate silicon and blue lines indicate subwavelength grating regions}
\label{fig:efield}
\end{figure}

Fig. \ref{fig:efield} shows the distribution of $E_{y}$ when the grating is illuminated by CW light at 1550 nm for the optimized configuration. This figure confirms that the optimized grating displays high unidirectionality, with almost no light being coupled into the right waveguide.

\section{Experimental validation}

Devices were fabricated on a 220 nm SOI platform in a CUMEC MPW run. To experimentally validate the proposed concept, we measured two test layouts. The first one was a 2-port structure consisting of a standard PDK grating coupler and the proposed ME-vGC (31 stripes), which was used to estimate the transmission of a single ME-vGC. The second one was a 3-port structure, where a short ME-vGC (8 stripes) was placed between two standard grating couplers, which was used to quantify unidirectionality by comparing the forward and backward transmission under identical alignment conditions. Five parameter variants were fabricated; below we report the best-performing device with $L_y=750$ nm, $L_x=450$ nm, $f_y=0.717$, $f_{\mathrm{SWG}}=0.500$, and $F_x=0.500$.

\begin{figure}[t]
  \centering
  \begin{minipage}[t]{0.48\linewidth}
    \centering
    \includegraphics[width=\linewidth]{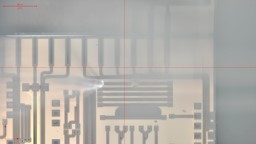}
    \vspace{2mm}
    {\small (a)}
  \end{minipage}\hfill
  \begin{minipage}[t]{0.48\linewidth}
    \centering
    \includegraphics[width=\linewidth]{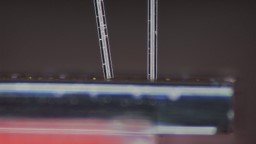}
    \vspace{2mm}
    {\small (b)}
  \end{minipage}
  \caption{Experimental setup for ME-vGC characterization on the probe station: (a) top view and (b) side view of the fiber alignment under microscope inspection.}
  \label{fig:exp_photo}
\end{figure}

Measurements were carried out on a probe station with fiber alignment controlled under an optical microscope. A depolarized ASE source was used for illumination, and the transmitted spectra were recorded using an Anritsu optical spectrum analyzer. In the 2-port layout, the transmission of a single ME-vGC was extracted by subtracting the contribution of one standard PDK grating coupler obtained from a reference GC--GC structure. In the 3-port layout, unidirectionality was defined as the difference between the forward and backward transmission, $\mathrm{FWD}-\mathrm{BWD}$, measured under the same alignment conditions. A photograph of the experimental setup is shown in Fig.~\ref{fig:exp_photo}.

\begin{figure}[t]
  \centering
  \includegraphics[width=0.7\linewidth]{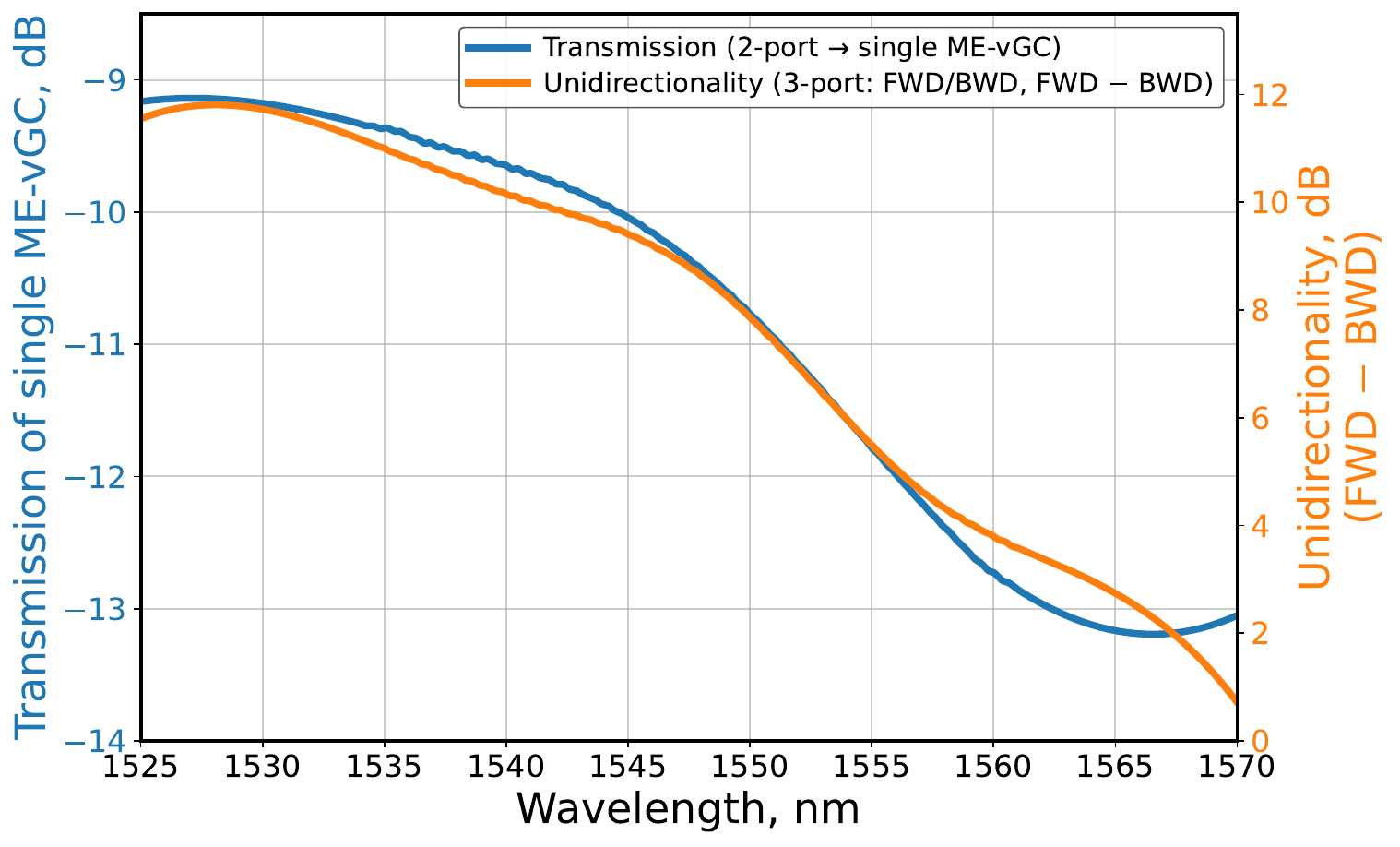}
  \caption{Measured transmission of a single ME-vGC (blue, left axis) and unidirectionality (orange, right axis) as a function of wavelength for the best-performing device.}
  \label{fig:exp_results}
\end{figure}

The measured spectra are shown in Fig.~\ref{fig:exp_results}. The best-performing device exhibits clear directional behavior, with a unidirectionality of $12.80\pm0.02$ dB near 1528 nm. The extracted transmission of a single ME-vGC is $-8.35\pm0.02$ dB. These results experimentally confirm the anisotropic coupling behavior predicted by simulation, although the measured insertion loss remains higher than the simulated value.

\section{Discussion}

The demonstrated design principle enables a relatively high perfectly vertical coupling efficiency of 41\% while retaining a fabrication-friendly single-etch grating geometry. The proposed approach can be further improved in several ways, including apodization of the structure, optimization of the grating “teeth” geometry, and implementation in a focusing configuration. In addition, the same design concept can be extended to grating couplers on other integrated photonics platforms, in particular, silicon nitride photonics. The proposed grating is promising for both fiber-to-chip and VCSEL-to-chip coupling, which are important for a broad range of integrated-photonics applications, including classical and quantum communications, LiDAR, and sensing.

At the same time, the current design is fundamentally constrained by the use of a single etching step and therefore is not expected to reach the efficiency of more advanced multi-etch grating couplers. In the experimental part, the results presented in this work should be regarded as preliminary. At present, only initial experimental data obtained from a single fabricated chip are available. Ongoing work includes measurements on multiple chips in order to collect sufficient statistics and evaluate reproducibility of the observed behavior. In parallel, the numerical modeling is being revisited to more accurately reflect the experimental conditions, and a deeper theoretical analysis of the coupling mechanism is being conducted. For this reason, the present manuscript is released in its current form as an arXiv preprint, while a more complete journal version will be prepared after the experimental dataset and theoretical analysis are finalized.

\section{Conclusion}
In this paper, we have presented a new bio-inspired concept for the design of perfectly vertical sub-wavelength grating couplers that combines fabrication simplicity with directional vertical coupling. Three-dimensional FDTD simulations were used to validate the concept and predict a coupling efficiency of 41\% at 1550 nm, a -1 dB bandwidth of about 50 nm, and unidirectionality of 20 dB at the design wavelength.

In addition to the numerical results, initial experimental measurements were carried out and provided a proof-of-concept demonstration of the proposed approach. In particular, anisotropic vertical coupling was experimentally observed, with measured unidirectionality reaching $12.80 \pm 0.02$ dB and extracted single-coupler transmission of $-8.35 \pm 0.02$ dB for the best-performing device. These results confirm the physical validity of the proposed design principle, although the present experimental dataset remains preliminary.

\paragraph{Disclosures}
The authors declare no conflicts of interest.
\paragraph{Data Availability Statement}
The simulation codes can be obtained from the authors upon request.

\bibliographystyle{unsrt}
\bibliography{sample}

@article{oskooi2010meep,
  title={MEEP: A flexible free-software package for electromagnetic simulations by the FDTD method},
  author={Oskooi, Ardavan F and Roundy, David and Ibanescu, Mihai and Bermel, Peter and Joannopoulos, John D and Johnson, Steven G},
  journal={Computer Physics Communications},
  volume={181},
  number={3},
  pages={687--702},
  year={2010},
  publisher={Elsevier}
}

@article{marchetti2019coupling,
  title={Coupling strategies for silicon photonics integrated chips},
  author={Marchetti, Riccardo and Lacava, Cosimo and Carroll, Lee and Gradkowski, Kamil and Minzioni, Paolo},
  journal={Photonics Research},
  volume={7},
  number={2},
  pages={201--239},
  year={2019},
  publisher={Optica Publishing Group}
}

@article{cheng2020grating,
  title={Grating couplers on silicon photonics: Design principles, emerging trends and practical issues},
  author={Cheng, Lirong and Mao, Simei and Li, Zhi and Han, Yaqi and Fu, HY},
  journal={Micromachines},
  volume={11},
  number={7},
  pages={666},
  year={2020},
  publisher={MDPI}
}

@article{vermeulen2018optical,
  title={Optical interfaces for silicon photonic circuits},
  author={Vermeulen, Diedrik and Poulton, Christopher V},
  journal={Proceedings of the IEEE},
  volume={106},
  number={12},
  pages={2270--2280},
  year={2018},
  publisher={IEEE}
}

@article{lu2016flip,
  title={Flip-chip integration of tilted VCSELs onto a silicon photonic integrated circuit},
  author={Lu, Huihui and Lee, Jun Su and Zhao, Yan and Scarcella, Carmelo and Cardile, Paolo and Daly, Aidan and Ortsiefer, Markus and Carroll, Lee and O’Brien, Peter},
  journal={Optics express},
  volume={24},
  number={15},
  pages={16258--16266},
  year={2016},
  publisher={Optica Publishing Group}
}

@article{han2016antireflective,
  title={Antireflective surface inspired from biology: A review},
  author={Han, ZW and Wang, Ze and Feng, XM and Li, Bo and Mu, ZZ and Zhang, JQ and Niu, SC and Ren, LQ},
  journal={Biosurface and Biotribology},
  volume={2},
  number={4},
  pages={137--150},
  year={2016},
  publisher={Elsevier}
}

@article{rytov1956electromagnetic,
  title={Electromagnetic properties of a finely stratified medium},
  author={Rytov, S},
  journal={Soviet Physics JEPT},
  volume={2},
  pages={466--475},
  year={1956}
}

@article{lalanne1998high,
  title={High-order effective-medium theory of subwavelength gratings in classical mounting: application to volume holograms},
  author={Lalanne, Philippe and Hugonin, Jean-Paul},
  journal={JOSA A},
  volume={15},
  number={7},
  pages={1843--1851},
  year={1998},
  publisher={Optical Society of America}
}

@article{lalanne1996effective,
  title={On the effective medium theory of subwavelength periodic structures},
  author={Lalanne, Philippe and Lemercier-Lalanne, Dominique},
  journal={Journal of Modern Optics},
  volume={43},
  number={10},
  pages={2063--2085},
  year={1996},
  publisher={Taylor \& Francis}
}

@article{cheben2018subwavelength,
  title={Subwavelength integrated photonics},
  author={Cheben, Pavel and Halir, Robert and Schmid, Jens H and Atwater, Harry A and Smith, David R},
  journal={Nature},
  volume={560},
  number={7720},
  pages={565--572},
  year={2018},
  publisher={Nature Publishing Group UK London}
}

@article{halir2015waveguide,
  title={Waveguide sub-wavelength structures: a review of principles and applications},
  author={Halir, Robert and Bock, Przemek J and Cheben, Pavel and Ortega-Mo{\~n}ux, Alejandro and Alonso-Ramos, Carlos and Schmid, Jens H and Lapointe, Jean and Xu, Dan-Xia and Wang{\"u}emert-P{\'e}rez, J Gonzalo and Molina-Fern{\'a}ndez, {\'I}{\~n}igo and others},
  journal={Laser \& Photonics Reviews},
  volume={9},
  number={1},
  pages={25--49},
  year={2015},
  publisher={Wiley Online Library}
}

\end{document}